\documentclass[a4paper]{article}
\pdfoutput=1
 
\usepackage{url}
\usepackage{xspace}

\usepackage{tlatex}

\title{\tlaplus\ Proofs\thanks{%
    This work was partially funded by INRIA-Microsoft Research Joint Centre, France.\newline
    It was presented at Dagstuhl seminar 12271 ``AI Meets Formal Software
    Development'' (July 2--6, 2012).
    A shorter version of this article appears as~\cite{cousineau:tla-proofs-fm}.}}

\author{
   Denis Cousineau\textsuperscript{1} \quad
   Damien Doligez\textsuperscript{2} \quad
   Leslie Lamport\textsuperscript{3} \\
   Stephan Merz\textsuperscript{4} \quad
   Daniel Ricketts\textsuperscript{5} \quad
   Hern\'an Vanzetto\textsuperscript{4}\\[2ex]
   {\small\textsuperscript{1}Inria - Universit\'e Paris Sud, Orsay, France}\\
   {\small\textsuperscript{2}Inria, Paris, France}\\
   {\small\textsuperscript{3}Microsoft Research, Mountain View, CA, U.S.A.}\\
   {\small\textsuperscript{4}Inria Nancy \& LORIA, Villers-l\`es-Nancy, France}\\
   {\small\textsuperscript{5}Department of Computer Science, University of California, San Diego, U.S.A.}
}

\date{}

\newcommand{\implies}{\Rightarrow}
\newcommand{\tlaplus}{\mbox{TLA\kern -.35ex$^+$}\xspace}
\newcommand{\tlatwo}{\mbox{TLA\kern -.35ex$^{+\mbox{\tiny 2}}$}\xspace}
\newcommand{\PM}{PM\xspace}

\newenvironment{noj}{\begin{array}[t]{@{}l@{}}}{\end{array}}
\newenvironment{conj}{\begin{array}[t]{@{\mbox{$\land\ $}}l}}{\end{array}}

\makeatletter
\newcommand{\step}[2]{{\tlatex \@pfstepnum{#1}{#2}}}
\makeatother

\def\S#1{\hspace*{#1em}}
\def\T#1{\hspace*{-#1pt}}

\makeatletter \let\str=\@w \makeatother

\newenvironment{display}{\begin{itemize}\item[]}{\end{itemize}}

\newcommand{\proofrule}[2]{\setlength{\arrayrulewidth}{.6pt}%
  {\ensuremath{\begin{array}[t]{@{}c@{}}%
       \begin{array}[t]{@{}l@{}} #1\raisebox{-.1em}{\strut}\end{array}\\
       \hline \raisebox{.1em}{\strut}#2\end{array}}}}

\begin{document}

\maketitle

\begin{abstract}
  \tlaplus is a specification language based on standard set theory and temporal
  logic that has constructs for hierarchical proofs. We describe how to write
  \tlaplus proofs and check them with TLAPS, the \tlaplus Proof System. We use
  Peterson's mutual exclusion algorithm as a simple example to describe the
  features of TLAPS and show how it and the Toolbox (an IDE for \tlaplus)
  help users to manage large, complex proofs.
\end{abstract}

\section{Introduction}

\tlaplus~\cite{lamport03tla} is a specification language originally designed for
specifying concurrent and distributed systems and their properties.
Specifications and properties are written as formulas of TLA, a linear-time
temporal logic. \tlaplus\ is based on TLA and Zermelo-Fraenkel set theory with
the axiom of choice; it also adds a module system for structuring
specifications. More recently, constructs for writing proofs have been added to
\tlaplus; these are derived from a hierarchical presentation of
natural-deduction proofs proposed for writing rigorous hand
proofs~\cite{lamport:howtoprove-21century}.

In this paper, we present the main ideas that guided the design of the proof
language and our experience with using the \tlaplus tools for verifying safety
properties of \tlaplus specifications. The \tlaplus Toolbox is an integrated
development environment (IDE) based on Eclipse for writing \tlaplus
specifications and running the \tlaplus tools on them, including the TLC model
checker and TLAPS, the \tlaplus proof system~\cite{chaudhuri:tlaps,tlaps}. In
particular, it provides 
commands to hide and unhide parts of a proof, allowing a user to focus on a
given proof step and its context. It is also invaluable to be able to run the model
checker on the same formulas that one reasons about.

The \tlaplus proof language and TLAPS have been designed to be independent of
any particular theorem prover. All interaction takes place at the level of
\tlaplus, letting the user focus on the specification of the algorithm being
developed. We do not expect users to have precise knowledge of the inner
workings of the back-end provers that TLAPS uses, although with experience users
learn about the strengths and weaknesses of the different provers---for example,
that SMT solvers excel at arithmetic.

TLAPS has a \emph{Proof Manager} (\PM) that transforms a proof into
individual proof obligations that it sends to back-end provers.
Currently, the main back-end provers are Isabelle/\tlaplus, an encoding
of \tlaplus as an object logic in Isabelle~\cite{wenzel:isabelle},
Zenon~\cite{bonichon07lpar}, a tableau prover for classical
first-order logic with equality, and a back-end for SMT solvers.
Isabelle serves as
the most trusted back-end prover, and when possible, we expect back-end provers to
produce a detailed proof that is checked by Isabelle. This is currently
implemented for the Zenon back-end, which can export its proofs as Isar
scripts that Isabelle can certify.

We explain how to write and check \tlaplus\ proofs,
using a tiny well-known example: a proof that Peterson's
algorithm~\cite{peterson:myths} implements
mutual exclusion. We start by writing the algorithm in
PlusCal~\cite{lamport:pluscal}, an algorithm language that is based on the
expression language of \tlaplus. The PlusCal code is translated to a \tlaplus
specification, which is what we reason about. Section~\ref{sec:proving}
introduces the salient features of the proof language and of TLAPS with the
proof of mutual exclusion. Liveness of Peterson's
algorithm (processes eventually enter their critical section) can also be
asserted and proved with \tlaplus. However, liveness reasoning makes full use of
temporal logic, and TLAPS cannot yet check temporal logic proofs. We therefore
discuss only mutual exclusion.

Section~\ref{sec:real-proofs} describes how \tlaplus, TLAPS, and the
Toolbox scale to realistic examples.  Their relation to other proof
systems is discussed in Section~\ref{sec:related}.  A concluding
section summarizes what we have done and our plans for future work.

\section{Modeling Peterson's Algorithm In \tlaplus}
\label{sec:peterson}

Peterson's algorithm 
is a classic, very simple
two-process mutual exclusion algorithm.  We specify the algorithm in \tlaplus and prove
that it satisfies mutual exclusion, meaning that no two processes are in their
critical sections at the same time.\footnote{The \tlaplus module containing the
  specification and proof is accessible at the TLAPS Web page~\cite{tlaps}.}

\subsection{From PlusCal To \tlaplus}
\label{sec:pluscal}

We will write Peterson's algorithm in the PlusCal algorithm language. To do so,
we have the Toolbox create an empty \tlaplus\ module. We name the two processes
$0$ and $1$, and we define an operator $Not$ so that $Not(0)=1$ and $Not(1)=0$:
\begin{display}
\begin{tlatex}
\@x{ Not ( i ) \.{\defeq}\, {\IF} i \.{=} 0 \.{\THEN} 1 \.{\ELSE} 0}%
\end{tlatex}
\end{display}
\begin{figure}[tb]
\newlength{\labelsdim}
\settowidth{\labelsdim}{$a3a$:}
\newcommand{\makelab}[1]{\makebox[\labelsdim][r]{$#1$: }}
\begin{tabbing}
\S{5}\=\+\texttt{-{}-}\textbf{algorithm} Peterson \{ \\
\S{1.5}\=\+
  \textbf{variables} $flag = [i \in \{0, 1\} \mapsto \FALSE]$, $turn = 0$;\\
  \textbf{process} $(proc \in \{0,1\})$ \{\\
  \S{1.5}\=\+ \makelab{a0} \textbf{while} $(\TRUE)$ \{ \\
     \makelab{a1}\S{1.5}\=   $flag[self] := \TRUE$; \\
     \makelab{a2}\>   $turn := Not(self)$; \\
     \makelab{a3a}\>  \textbf{if} $(flag[Not(self)])$
                      \{\textbf{goto} $a3b$\} \textbf{else} 
                      \{\textbf{goto} $cs$\} ; \\
     \makelab{a3b}\>  \textbf{if} $(turn = Not(self))$ 
                      \{\textbf{goto} $a3a$\} \textbf{else} 
                      \{\textbf{goto} $cs$\} ; \\
     \makelab{cs}\>   \textbf{skip};  $\backslash*$ critical section \\
     \makelab{a4}\>   $flag[self] := \FALSE$; \\
     \hspace*{\labelsdim}\ \}  $\backslash*$  end while \- \\
    \} $\backslash*$  end process \- \\
  \S{.5}\} $\backslash*$  end algorithm 
\end{tabbing}
\caption{Peterson's algorithm in PlusCal.}
\label{fig:the-algorithm}
\end{figure}
The PlusCal code for Peterson's algorithm is shown in
Figure~\ref{fig:the-algorithm}; it appears in a comment in the \tlaplus\ module.%
\footnote{The figure shows the pretty-printed version of PlusCal code
  and \tlaplus formulas.  As an example of how they are typed,
  here is the \textsc{ascii} version of the \textbf{variables}
  declaration:\quad
  \texttt{variables flag = [i \textbackslash in \{0, 1\} |-> FALSE], turn = 0;}
}
The \textbf{variables} statement declares the variables and their initial
values. For example, the initial value of $flag$ is an array such that
$flag[0]=flag[1]=\FALSE$. (Mathematically, an array is a function; the
\tlaplus notation $[x \in S \mapsto e]$ for writing functions 
is similar to a lambda expression.)
To specify a multiprocess
algorithm, it is necessary to specify what its atomic actions are.  In
PlusCal, an atomic action consists of the execution from one label to the
next.  With this brief explanation, the reader should be able to
figure out what the code means.

\begin{figure}[tp]
\begin{tlatex}

\@xx{}%
\@x{ {\VARIABLES} flag ,\, turn ,\, pc}%
\@x{ vars \.{\defeq} {\langle} flag ,\, turn ,\, pc {\rangle}}%
\par\vspace{6.01pt}%
 \@x{ Init\@s{2.02} \.{\defeq} \.{\land} flag\@s{2.82} \.{=} [ i \.{\in} \{ 0
 ,\, 1 \} \.{\mapsto} {\FALSE} ]}%
\@x{\T{2.82}\@s{37.72} \.{\land} turn \.{=} 0}%
 \@x{\T{2.82}\@s{37.72} \.{\land} pc \.{=} [ self \.{\in} \{ 0 ,\, 1 \}
 \.{\mapsto}\@w{a0} ]}%
\par\vspace{6.01pt}%
\@x{ a0 ( self ) \.{\defeq} \.{\land} pc [ self ] \.{=}\@w{a0}}%
 \@x{\T{4.04}\@s{53.97} \.{\land} pc \.{'} \.{=} [ pc {\EXCEPT} {\bang} [ self ]
 \.{=}\@w{a1} ]}%
\@x{\T{4.04}\@s{53.97} \.{\land} {\UNCHANGED} {\langle} flag ,\, turn {\rangle}}%
\par\vspace{6.01pt}%
\@x{ a1 ( self ) \.{\defeq} \.{\land} pc [ self ] \.{=}\@w{a1}}%
 \@x{\T{4.04}\@s{53.97} \.{\land} flag \.{'} \.{=} [ flag {\EXCEPT} {\bang} [ self ]
 \.{=} {\TRUE} ]}%
 \@x{\T{4.04}\@s{53.97} \.{\land} pc \.{'} \.{=} [ pc {\EXCEPT} {\bang} [ self ]
 \.{=}\@w{a2} ]}%
\@x{\T{4.04}\@s{53.97} \.{\land} turn \.{'} \.{=} turn}%
\par\vspace{6.01pt}%
\@x{ a2 ( self ) \.{\defeq} \.{\land} pc [ self ] \.{=}\@w{a2}}%
\@x{\T{4.04}\@s{53.97} \.{\land} turn \.{'} \.{=} Not ( self )}%
 \@x{\T{4.04}\@s{53.97} \.{\land} pc \.{'} \.{=} [ pc {\EXCEPT} {\bang} [ self ]
 \.{=}\@w{a3a} ]}%
\@x{\T{4.04}\@s{53.97} \.{\land} flag \.{'} \.{=} flag}%
\par\vspace{6.01pt}%
\@x{ a3a ( self ) \.{\defeq} \.{\land} pc [ self ] \.{=}\@w{a3a}}%
\@x{\T{4.48}\@s{59.85} \.{\land} {\IF} flag [ Not ( self ) ]}%
 \@x{\T{12}\@s{83.11} \.{\THEN} pc \.{'} \.{=} [ pc {\EXCEPT} {\bang} [ self ]
 \.{=}\@w{a3b} ]}%
 \@x{\T{12}\@s{83.11} \.{\ELSE} pc \.{'} \.{=} [ pc {\EXCEPT} {\bang} [ self ]
 \.{=}\@w{cs} ]}%
\@x{\T{4.48}\@s{59.85} \.{\land} {\UNCHANGED} {\langle} flag ,\, turn {\rangle}}%
\par\vspace{6.01pt}%
\@x{ a3b ( self )\@s{0.64} \.{\defeq} \.{\land} pc [ self ] \.{=}\@w{a3b}}%
\@x{\T{4.48}\@s{59.85} \.{\land} {\IF} turn \.{=} Not ( self )}%
 \@x{\T{12}\@s{83.11} \.{\THEN} pc \.{'} \.{=} [ pc {\EXCEPT} {\bang} [ self ]
 \.{=}\@w{a3a} ]}%
 \@x{\T{12}\@s{83.11} \.{\ELSE} pc \.{'} \.{=} [ pc {\EXCEPT} {\bang} [ self ]
 \.{=}\@w{cs} ]}%
\@x{\T{4.48}\@s{59.85} \.{\land} {\UNCHANGED} {\langle} flag ,\, turn {\rangle}}%
\par\vspace{6.01pt}%
\@x{ cs ( self )\@s{1.36} \.{\defeq} \.{\land} pc [ self ] \.{=}\@w{cs}}%
 \@x{\T{4.04}\@s{53.97} \.{\land} pc \.{'} \.{=} [ pc {\EXCEPT} {\bang} [ self ]
 \.{=}\@w{a4} ]}%
\@x{\T{4.04}\@s{53.97} \.{\land} {\UNCHANGED} {\langle} flag ,\, turn {\rangle}}%
\par\vspace{6.01pt}%
\@x{ a4 ( self ) \.{\defeq} \.{\land} pc [ self ] \.{=}\@w{a4}}%
 \@x{\T{4.04}\@s{53.97} \.{\land} flag \.{'} \.{=} [ flag {\EXCEPT} {\bang} [ self ]
 \.{=} {\FALSE} ]}%
 \@x{\T{4.04}\@s{53.97} \.{\land} pc \.{'} \.{=} [ pc {\EXCEPT} {\bang} [ self ]
 \.{=}\@w{a0} ]}%
\@x{\T{4.04}\@s{53.97} \.{\land} turn \.{'} \.{=} turn}%
\par\vspace{6.01pt}%
 \@x{ proc ( self ) \.{\defeq} a0 ( self ) \.{\lor} a1 ( self ) \.{\lor} a2 (
 self ) \.{\lor} a3a ( self ) \.{\lor} a3b ( self )}%
\@x{\@s{76.44} \.{\lor} cs ( self ) \.{\lor} a4 ( self )}%
\par\vspace{6.01pt}%
\@x{ Next \.{\defeq} \E\, self \.{\in} \{ 0 ,\, 1 \} \.{:} proc ( self )}%
\par\vspace{6.01pt}%
\@x{ Spec\@s{1.46} \.{\defeq} Init \.{\land} {\Box} [ Next ]_{ vars}}%

\end{tlatex}
\caption{A pretty-printed version of the \tlaplus translation, slightly simplified.}
\label{fig:translation}
\end{figure}

The PlusCal translator, accessible through a Toolbox menu, generates a \tlaplus
specification from the PlusCal code of the algorithm.
Figure~\ref{fig:translation} gives the generated \tlaplus translation.%
\footnote{For clarity of presentation, we have simplified the translation
  slightly by ``in-lining'' a definition. The proof we develop works for the
  unmodified translation if we add a global declaration that causes the
  definition to be expanded throughout the proof.}
The PlusCal compiler adds a variable
$pc$, which explicitly records the control state of each process. For example,
control in process $i$ is at $cs$ iff $pc[i]$ equals the string \str{cs}.

The heart of the \tlaplus specification consists of the initial predicate
$Init$, which describes the initial state, and the next-state relation $Next$,
which describes how the state can change. 
Given the PlusCal code, the meaning of formula $Init$ in the figure is
straightforward.  The formula $Next$ is a predicate on
old-state/new-state pairs.  Unprimed variables refer to the old state
and primed variables to the new state.  Formula $Next$ is the
disjunction of the two formulas $proc(0)$ and $proc(1)$, and each
$proc(self)$ is the disjunction of seven formulas---one for each label
in the body of the \textbf{process}.  The formula $a0(self)$ specifies
the state change performed by process $self$ executing an atomic
action starting at label $a0$, and similarly for the other six labels.
(If $f$ is a function, the \tlaplus notation $[f \EXCEPT ![arg] = exp]$ denotes
the function that is equal to $f$ except that it maps $arg$ to $exp$.)
The reader should be able to figure out the meaning of the \tlaplus
notation and of formula $Next$ by comparing these seven definitions
with the corresponding PlusCal code.

The temporal formula $Spec$ is the complete specification.  It is
satisfied by a behavior (i.e., an $\omega$-sequence of states) 
iff the behavior starts in a state satisfying
$Init$ and each of its steps (pairs of successive states) either
satisfies $Next$ or else leaves the values of the three variables
$flag$, $turn$, and $pc$ unchanged.\footnote{``Stuttering steps'' that
  leave all variables unchanged are allowed in order to make refinement
  simple~\cite{lamport:what-good}.} The $\Box$ is the ordinary
\emph{always} operator of linear-time temporal logic, and
$[Next]_{vars}$ is an abbreviation for \,\mbox{$Next \,\lor\,
\UNCHANGED vars$}\,, where $\UNCHANGED vars$ is an abbreviation for
$vars'=vars$ and priming an expression means priming all the variables
that occur in it.

\subsection{Validation Through Model Checking}
\label{sec:model-checking}

Before trying to prove that the algorithm is correct, we use TLC,
the \tlaplus model checker, to check it for errors.
We first instruct the Toolbox to have TLC check for
``execution errors''.\footnote{The translation is a
  temporal logic formula, so there is no obvious definition of an
   execution error.  An execution error occurs in a
  behavior if whether or not the behavior satisfies the formula is not
  specified by the semantics of \tlaplus---for example, because
  the semantics do not specify whether or not 0 equals $\FALSE$.}
What are type errors in typed languages are one source of execution
errors in \tlaplus.

The Toolbox runs TLC on a model of a \tlaplus specification.  A model
usually assigns particular values to specification constants, such as
the number $N$ of processes. It can also restrict the set of states
explored, which is useful if the specification allows an infinite number
of reachable states.  For this trivial example, there are no constants
to specify and only 58 reachable states.  TLC finds no execution
errors.

We next check if the algorithm actually satisfies mutual exclusion.
Since we made execution of the critical section an atomic action,
mutual exclusion means that the two processes never both have control
at label $cs$.  Mutual exclusion therefore holds iff the
following predicate $MutualExclusion$ is an invariant of the
algorithm---meaning that it is true in all reachable states:
\begin{display}
\begin{tlatex}
 \@x{ MutualExclusion \.{\defeq}\@s{4.1} ( pc [ 0 ] \.{\neq}\@w{cs} ) \.{\lor}
 ( pc [ 1 ] \.{\neq}\@w{cs} )}%
\end{tlatex}
\end{display}
TLC reports that the algorithm indeed satisfies this invariant. Peterson's
algorithm is so simple that TLC has checked that all possible
executions satisfy mutual exclusion. For more interesting algorithms that have
an infinite set of reachable states, TLC is no longer able to exhaustively
verify all executions, and correctness must be proved deductively. Still, TLC is
invaluable for catching errors in the algorithm or its formal model: the
effort required for running TLC is incomparably lower than that for writing a
formal proof.

\section{Proving Mutual Exclusion For Peterson's Algorithm}
\label{sec:proving}

We now describe a deductive correctness proof of Peterson's
algorithm in \tlaplus. Proofs of more interesting algorithms follow the same
basic structure, but they are longer.
Section~\ref{sec:real-proofs} describes how \tlaplus\ proofs scale to larger
algorithms.

\subsection{The High-Level Proof}
\label{sec:high-level-proof}

The assertion that Peterson's algorithm implements mutual exclusion
is formalized in \tlaplus\ as:
 \[ \THEOREM Spec \implies \Box MutualExclusion
 \]
The standard method of proving this invariance property is to find an
inductive invariant $Inv$ that implies $MutualExclusion$.  An
inductive invariant is one that is true in the initial state and whose
truth is preserved by the next-state relation.  \tlaplus\ proofs are
hierarchically structured and are generally written top-down.  The
top level of this invariance proof is shown in
Figure~\ref{fig:high-level-pf}.  Step \step{1}{2} asserts that 
the truth of $Inv$ is preserved by the next-state relation.

\begin{figure}[t]
\begin{tlatex}
\@x{ {\THEOREM} Spec \.{\implies} {\Box} MutualExclusion}%
\@x{\@pfstepnum{1}{1.}\  Init \.{\implies} Inv}%
 \@x{\@pfstepnum{1}{2.}\  Inv \.{\land} [ Next ]_{ vars} \.{\implies} Inv
 \.{'}}%
\@x{\@pfstepnum{1}{3.}\  Inv \.{\implies} MutualExclusion}%
\@x{\@pfstepnum{1}{4.}\  {\QED}}%
\end{tlatex}
\caption{The high-level proof.}
\label{fig:high-level-pf}
\end{figure}

Each proof in the hierarchy ends with a \QED step that asserts the
goal of that proof, the \QED step for the top level asserting the
statement of the theorem.  We usually write the \QED step's proof
first.  This \QED step follows easily from \step{1}{1}, \step{1}{2},
and \step{1}{3} by propositional logic and the following two
temporal-logic proof rules:
 \[ \proofrule{I \land [N]_v \implies I'}{I \land \Box[N]_v \implies \Box I}
   \S{4}
    \proofrule{P \implies Q}{\Box P \implies \Box Q}
 \] 
However, TLAPS does not yet handle temporal reasoning, so we omit the
proof of the \QED step.  When temporal reasoning is added to TLAPS, we expect it easily
to check such a trivial proof.

\begin{figure}[b]
\(\begin{noj}
  TypeOK \defeq
  \begin{conj}
    pc \in [\:\{0,1\} \rightarrow \{\str{a0}, \str{a1}, \str{a2}, \str{a3a}, \str{a3b}, \str{cs}, \str{a4}\}\:]\\
    turn \in \{0,1\}\\
    flag \in [\:\{0,1\} \rightarrow \BOOLEAN]
  \end{conj}\\[27pt]
  I \defeq \forall i \in \{0,1\}:\\
  \hspace*{3.8em}\begin{conj}
    pc[i] \in \{\str{a2}, \str{a3a}, \str{a3b}, \str{cs}, \str{a4}\} \implies flag[i]\\
    pc[i] \in \{\str{cs}, \str{a4}\} \implies
    \begin{conj}
      pc[Not(i)] \notin \{\str{cs}, \str{a4}\}\\
      pc[Not(i)] \in \{\str{a3a}, \str{a3b}\} \implies turn = i
    \end{conj}
  \end{conj}\\[27pt]
  Inv \defeq TypeOK \land I
\end{noj}\)
\caption{The inductive invariant.}
\label{fig:inv}
\end{figure}

To continue the proof, we must define the inductive invariant $Inv$.
(A definition must precede its use, so the definition of $Inv$ appears
in the module before the proof.) 
Figure~\ref{fig:inv} defines $Inv$ to be the conjunction of two formulas.
The first, $TypeOK$, is a ``type-correctness'' invariant, asserting that the
values of all variables are elements of the expected sets.  (The
expression $[S\rightarrow T]$ is the set of all functions whose domain is
$S$ and whose range is a subset of $T$.)  In an untyped logic like that of
\tlaplus, almost any inductive invariant must assert type correctness.
The second conjunct, $I$, is the interesting one that explains why
Peterson's algorithm implements mutual exclusion.

There is no point trying to prove that a formula is an inductive
invariant if TLC can show that it's not even an invariant.  So, we
first run TLC to test if $Inv$ is an invariant. In the simple case of
Peterson's algorithm, TLC can check not only that it is an invariant, but
that it is an inductive invariant.  We check that $Inv$ is an
inductive invariant of $Spec$ by checking that it is an (ordinary)
invariant of the specification 
 \,\mbox{$Inv \land \Box[Next]_{vars}$}\,,
obtained from $Spec$ by replacing the initial condition by $Inv$.
In most real examples, TLC can at best check an inductive invariant on
a tiny model---one that is too small to gain any confidence that
it really is an inductive invariant.  However, TLC can still often find
simple errors in an inductive invariant.

\subsection{Leaf Proofs for Steps \step{1}{1}--\step{1}{3}}
\label{sec:leaf-proofs}

We now prove steps \step{1}{1}--\step{1}{3}.  We can prove them in any
order; let us start with \step{1}{1}.  We expect this step to follow
easily from the definitions of $Init$ and $Inv$ and simple properties
of sets and functions.  TLAPS knows about sets and functions, but it
does not expand definitions unless directed to do so.  (In complex
proofs, automatically expanding definitions often leads to formulas
that are too big for provers to handle.)  We assert that the
step follows from simple math and the definitions of $Init$ and $Inv$
by writing the following leaf proof immediately after the step:
\begin{display}
\textsc{by def} $Init$, $Inv$
\end{display}
We then tell the Toolbox to run TLAPS to check this proof.  It does so
and reports that the prover failed to prove the following obligation:
\begin{display}
\small
\begin{verbatim}
ASSUME NEW VARIABLE flag,
       NEW VARIABLE turn,
       NEW VARIABLE pc
PROVE  (/\ flag  =  [i \in {0, 1} |-> FALSE]
        /\ turn  =  0
        /\ pc  =  [self \in {0, 1} |-> "a0"])
        =>  TypeOK  /\  I
\end{verbatim}
\end{display}
This obligation is exactly what TLAPS's back-end provers are trying to
prove.  They are given no other facts.  In particular, the
provers know nothing about $TypeOK$ and $I$, so they obviously can't
prove the obligation.  We have to tell TLAPS also to use the definitions
of $TypeOK$ and $I$.  We do that by making the obvious change to the
\textsc{by} proof, after which TLAPS easily proves the step.
Forgetting to expand some definitions is a common mistake, and looking at the
formula displayed by the Toolbox usually reveals which definitions need to be
invoked.

Step \step{1}{3} is proved the same way, by simply expanding the
definitions of $MutualExclusion$, $Inv$, $I$, and $Not$.  We
next try the same technique on \step{1}{2}.  A little thought shows
that we have to tell TLAPS to expand all the definitions in the module
up to and including the definition of $Next$, except for the
definition of $Init$.
However,
when we direct TLAPS to prove the step, it fails to do so, reporting a
65-line proof obligation. 

TLAPS uses Zenon and Isabelle as its default back-end provers, first
trying Zenon and then trying Isabelle if Zenon fails to find a proof.
However, TLAPS also includes an SMT solver back-end~\cite{merz:smt-tlaps}
that is capable of
handling larger ``shallow'' proof obligations---in particular, ones
that do not contain significant quantifier reasoning.  We instruct
TLAPS to use the SMT back-end when proving the current step by writing
\begin{display}
\textsc{by} SMT \textsc{def} \ldots
\end{display}
The SMT back-end translates the proof obligation to
SMT-LIB~\cite{smtlib}, the standard input language for different SMT
solvers, and calls an SMT solver (CVC3 by default) to try to prove the
resulting formula.  CVC3 proves step \step{1}{2} in a 
few seconds.
Variants of the SMT back-end translate to the native input languages of
Yices and Z3, which sometimes perform better than does CVC3 using the
standard SMT-LIB translation.

\subsection{A Hierarchical Proof of Step \step{1}{2}}
\label{sec:hierarchical-proof}

For sufficiently complicated examples, an SMT solver will not be able
to prove inductive invariance as a single obligation.  The proof will
have to be hierarchically decomposed.  To illustrate how this is done,
we now write a proof of \step{1}{2} that can be checked using only the
Zenon and Isabelle back-end provers.  

\begin{figure}[bt]
\begin{tlatex}
 \@x{\@pfstepnum{1}{2.}\  Inv \.{\land} [ Next ]_{ vars} \.{\implies} Inv
 \.{'}}%
\@x{\@s{8.2}\@pfstepnum{2}{1.}\  {\SUFFICES} {\ASSUME} Inv ,\, Next}%
\@x{\T{5.7}\@s{76.37} {\PROVE} Inv \.{'}}%
\@x{\@s{8.2}\@pfstepnum{2}{2.}\  TypeOK \.{'}}%
\@x{\@s{8.2}\@pfstepnum{2}{3.}\  I \.{'}}%
\@x{\@s{8.2}\@pfstepnum{2}{4.}\  {\QED}}%
\end{tlatex}
\caption{The top-level proof of \step{1}{2}.}
\label{fig:1-2pf}
\end{figure}

Step \step{1}{2} and its top-level proof appear in
Figure~\ref{fig:1-2pf}.  The first step in the proof of an implication
like this would normally be:
\begin{display}
$\begin{array}{lll}
 \step{2}{1}.\ \SUFFICES & \ASSUME & Inv,\ [Next]_{vars} \\
                        & \PROVE & Inv'
 \end{array}
$
\end{display}
This step asserts that to prove the current goal, which is step
\step{1}{2}, it suffices to assume that $Inv$ and $[Next]_{vars}$ are
true and prove $Inv'$.  The step also changes the goal of the rest of
the level-2 proof to $Inv'$ and allows the assumptions $Inv$ and
$[Next]_{vars}$ to be used in the rest of the proof.  This step's
assertion is obviously true, and TLAPS will check the one-word leaf
proof \textsc{obvious}.  However, the proof of Figure~\ref{fig:1-2pf}
does something a little different.

Since the assumption $[Next]_{vars}$ equals 
  \,\mbox{$Next \lor \UNCHANGED vars$}\,,
it leaves two cases to be proved: (i)~$Next$ is true and (ii)~all
variables are unchanged, so their primed values equal their unprimed
values.  The proof in the second case is trivial, and TLAPS should
have no trouble checking it.  In Figure~\ref{fig:1-2pf}, the
assumption in the \textsc{suffices} statement is $Next$ rather than
$[Next]_{vars}$, so the remainder of the proof only has to consider
case~(i).  To show that it suffices to prove $Inv'$ under this
stronger assumption, the proof of that \textsc{suffices} step has to
prove case~(ii).

\begin{sloppypar}
The remainder of the level-2 proof is straightforward.  Since $Inv$
equals \,\mbox{$TypeOK \land I$}, the goal $Inv'$ is the conjunction
of the two formulas $TypeOK'$ and $I'$.  We therefore decompose the
proof by proving each conjunct separately.  
The proof of the \QED step is simply
\end{sloppypar}
\begin{display}
\begin{tlatex}
\@x{{\BY}\@pfstepnum{2}{2} ,\,\@pfstepnum{2}{3}\  {\DEF} Inv}%
\end{tlatex}
\end{display}
Observe that we have to tell TLAPS exactly what facts to use as well
as what definitions to expand.  

We next prove \step{2}{1}--\step{2}{3}.  Zenon proves \step{2}{1} when
the definitions of $vars$, $Inv$, $TypeOK$, and $I$ are expanded.
Note that the definition of $Next$ is not needed.  To prove
\step{2}{2} and \step{2}{3}, we need to use the definition of
$Next$---that is, with all definitions expanded down to \tlaplus\
primitives---as well as the definition of $Inv$.  We also have to use
the assumption that $Inv$ and $Next$ are true, introduced by step
\step{2}{1}.  This leads us to try the following proof for
\step{2}{2}.
\begin{display}
\begin{tlatex}
 \@x{{\BY}\@pfstepnum{2}{1}\  {\DEFS} Inv ,\, TypeOK ,\, Next ,\,
 proc ,\, a0 ,\, a1 ,\, a2 ,\, a3a ,\, a3b ,\, cs ,\, a4 ,\, Not}
\end{tlatex}
\end{display}
Instead of the reference to step \step{2}{1} in the \BY clause, we could also
name the required facts directly and write
\begin{display}
  $\BY Inv, Next\ \DEFS \ldots$
\end{display}
The proof manager checks that $Inv$ and $Next$ indeed follow from the currently
available assumptions.

Zenon fails on this proof, but Isabelle succeeds.  However, both Zenon
and Isabelle fail on the corresponding proof of \step{2}{3} (which
requires also using the definition of $I$).  To prove it (with only
Zenon and Isabelle), we need one more level of proof.  That level
appears in Figure~\ref{fig:complete-proof}, which contains the
complete proof of the theorem.

\begin{figure}[tb]
  \(\begin{noj}
    \THEOREM Spec \implies \Box MutualExclusion\\
    \step{1}{1.}\ Init \implies Inv\\
    \quad\BY \DEFS Init,\, Inv,\, TypeOK,\, I\\
     \step{1}{2.}\ Inv \land [Next]_{vars} \implies Inv'\\
     \quad\begin{noj}
       \step{2}{1.}\ \SUFFICES
         \begin{noj}
           \ASSUME Inv,\, Next\\
           \PROVE Inv'
         \end{noj}\\
       \quad\BY \DEFS Inv,\, TypeOK,\, I,\, vars\\
       \step{2}{2.}\ TypeOK'\\
       \quad\BY \step{2}{1}\ \DEFS Inv,\, TypeOK,\, Next,\, proc,\, a0,\, a1,\, a2,\, a3a,\, a3b,\, cs,\, a4,\, Not\\
       \step{2}{3.}\ I'\\
       \quad\begin{noj}
         \step{3}{1.}\ \SUFFICES
           \begin{noj}
             \ASSUME\ \NEW\ j \in \{0,1\}\\
             \PROVE\ I!(j)'
           \end{noj}\\
         \quad\BY \DEF I\\
         \step{3}{2.}\ \PICK\ i \in \{0,1\}: proc(i)\\
         \quad\BY \step{2}{1}\ \DEF Next\\
         \step{3}{3.}\ \CASE i=j\\
         \quad\BY
           \begin{noj}
             \step{2}{1},\, \step{3}{2},\, \step{3}{3}\\
             \DEFS Inv,\, I,\, TypeOK,\, proc,\, a0,\, a1,\, a2,\, a3a,\, a3b,\, cs,\, a4,\, Not
           \end{noj}\\
         \step{3}{4.}\ \CASE i \neq j\\
         \quad\BY
           \begin{noj}
             \step{2}{1},\, \step{3}{2},\, \step{3}{4}\\
             \DEFS Inv,\, I,\, TypeOK,\, proc,\, a0,\, a1,\, a2,\, a3a,\, a3b,\, cs,\, a4,\, Not
           \end{noj}\\
         \step{3}{5.}\ \QED\\
         \quad\BY \step{3}{3},\, \step{3}{4}
       \end{noj}\\
       \step{2}{4.}\ \QED\\
       \quad\BY \step{2}{2},\, \step{2}{3}\ \DEF Inv
     \end{noj}\\
     \step{1}{3.}\ Inv \implies MutualExclusion\\
     \quad\BY\ \DEF MutualExclusion, Inv, I, Not\\
     \step{1}{4.}\ \QED\\
     \quad\PROOF \OMITTED
  \end{noj}\)

\caption{The complete hierarchical proof.}
\label{fig:complete-proof}
\end{figure}

Since priming a formula means priming all variables in it, the goal
$I'$ has the form $\A i \in \{0,1\}: exp(i)'$.  A standard way to prove
this formula is by 
$\A$-introduction: we introduce a new variable, say $j$, we assume
$j\in\{0,1\}$, and we prove $exp(j)'$.  
%
\tlaplus\ provides a notation for naming subexpressions of a
definition.  With that notation, the expression $exp(j)$ is written
$I!(j)$.  This leads us to begin the proof of \step{2}{3} with the
\textsc{suffices} step \step{3}{1} of Figure~\ref{fig:complete-proof}
and its simple proof.

The assumption $Next$ (introduced by \step{2}{1}) equals
 \,\mbox{$\E self \in \{0,1\}: proc(self)$}\,.
A standard way to use such an assumption is by $\E$-elimination: we
pick some value of $self$ such that $proc(self)$ is true.  That is
what step \step{3}{2} does, naming the value~$i$.

We simplified our task to proving $I!(j)'$ instead of $I'$, using
$proc(i)$ instead of $Next$, which eliminates two quantifiers.
However, Zenon and Isabelle still cannot prove the goal in a single
step. The usual way to decompose the proof that process $i$ preserves an
invariant is to
show that each separate atomic action of process $i$ preserves the
invariant.  In mathematical terms, $proc(i)$ is the disjunction of the
seven formulas $a0(i)$, \ldots, $a4(i)$, each describing one of the
process's atomic action.  We can decompose the proof by considering
each of the seven formulas as a separate case.

While this is the usual procedure, Peterson's algorithm is simple
enough that it is not necessary.  Instead, we just have to help the
back-end provers by splitting the proof into the two cases of $i=j$
and $i\neq j$.  The reader can see how this is done in
Figure~\ref{fig:complete-proof}.  Observe that in the proof of
\textsc{case} statement \step{3}{3}, the name \step{3}{3} refers to
the case assumption $i=j$.  There is no explicit use of \step{3}{1}
because a \textsc{new} assumption in an \textsc{assume} is used by
default in all proofs in the assumption's scope.  The same is true of
the formula $i\in\{0,1\}$ asserted by the \textsc{pick} step.  (This
is a pragmatic choice in the design of TLAPS, based on the observation
that such facts are used so often.)

\section{Writing Real Proofs}
\label{sec:real-proofs}

We have described how one writes and checks a \tlaplus proof of a
tiny example.  
Several larger case studies have been carried out using the system.
These include verifications of Byzantine
Paxos~\cite{lamport:byzantine-paxos}, the Memoir security
architecture~\cite{parno:memoir}, and the lookup and join protocols
of the Pastry algorithm for maintaining a distributed hashtable over a
peer-to-peer network~\cite{lu:pastry}.
\tlaplus and TLAPS, with its Toolbox interface,
provide a number of features that help manage the complexity of large
proofs.

\subsection{Hierarchical Proofs And The Proof Manager}
\label{sec:hierarchy}

The most important aid in writing large proofs is \tlaplus's
hierarchical and declarative proof language, where intermediate proof
obligations are stated explicitly.  While declarative proofs are more
verbose than standard tactic scripts, they are easier to understand
and maintain because the information on what is currently being proved
is available at each point.  Hierarchical proofs enable a user to keep
decomposing a complex proof into smaller steps until the steps become
provable by one of the back-end provers.

In logical terms, proof steps correspond to natural-deduction sequents whose
validity must be established in the current context (containing constant and
variable symbols, assumptions, and already-established facts). The Proof Manager
tracks the context, which is modified by non-leaf proof steps. For leaf proof
steps, it sends the corresponding sequent to the back-end
provers, and records the status of the step's proof (succeeded, failed, canceled
by the user, or omitted).

Because proof obligations are independent of one another,
users can develop proofs in any order and work on the proof of a step
independently of the state of the proof of other steps.  This
permits them to concentrate on the part of a planned proof that is most likely
to be wrong and require changes to other parts.  The
Toolbox makes it easy to instruct TLAPS to check the proof of
everything in a file, of any single theorem, or of any single step.
It displays every obligation whose proof fails or
is taking too long; in the latter case the user can cancel the proof.
Clicking on the obligation shows the part of the proof that generated
it.

A linear presentation, as in Figure~\ref{fig:complete-proof}, is
unsuitable for reading or writing large proofs.  The Toolbox's editor helps
reading and writing large \tlaplus proofs, providing commands that
show or hide particular subproofs.  Commands to hide a proof
or view just its top level aid in reading a proof.  A
command that is particularly useful when writing a subproof is one
that hides all preceding steps that cannot be used in that subproof
because of their positions in the hierarchy.

\tlaplus's hierarchical proofs provide a much more powerful mechanism
for structuring complex proofs than the conventional approach using
lemmas.  In a \tlaplus\ proof, each step with a non-leaf proof is
effectively a lemma.  One typical 1100-line invariance
proof~\cite{lamport:byzantine-paxos}
contains 100 such steps.  A conventional linear proof with 100 lemmas
would be impossible to read.

\subsection{Fingerprinting: Tracking The Status Of Proof Obligations}
\label{sec:fingerprinting}

During proof development, a user repeatedly modifies the proof structure or changes
details of the specification. Rerunning the back-end provers on a sizable proof
takes time. 
By default, TLAPS does not re-prove an obligation that it has already
proved---even if the proof has been reorganized and the step that generated it
has been moved, or if the step was removed from the proof and reinserted in a
later version.  
It can also show the user the impact of a change by
indicating which parts of the existing proof must be re-proved.

The Proof Manager computes a \emph{fingerprint} of every obligation, 
which it
stores, along with the obligation's status, in a separate file. Technically, a proof
obligation is canonically represented as a lambda term, 
with bound variables
replaced by de Bruijn indices~\cite{deBruijn72} such that their actual names in the
\tlaplus proof are irrelevant. The context is minimized by erasing symbols and
hypotheses that are not used in the step. The fingerprint is a compact representation
of the resulting term, which is therefore insensitive
to structural modifications of the proof context that do not affect the obligation's logical
validity.

The Toolbox displays the proof status of each step, indicating by
color whether the step has been proved or some obligation in its proof
has failed or has been omitted.
Looking up an obligation's status takes little time, so the user can tell TLAPS
to re-prove a step or a theorem even if only a small part of the proof has
changed; TLAPS will recognize any obligation that has not changed and
will not attempt to prove it anew.
There is also a check-status command that displays the proof status without
actually launching any proofs. 

An incident that occurred in the Byzantine Paxos proof reveals the
advantages of our method of writing proofs.  The third author wrote
the safety proof primarily as a way of debugging TLAPS, spending a
total of several weeks over several months on it.  Later, when writing a
paper about the algorithm, he discovered that it did not satisfy the
desired liveness property, so it had to be modified.  He changed the
algorithm, fixed minor bugs found by TLC, and reproved the safety
property---all in a day and a half, with about 12 hours of
actual work.  He was able to do it that fast because of the
hierarchical proof structure, TLAPS's fingerprinting mechanism (about
3/4 of the proof obligations in the new proof had already been
proved), and the Toolbox's aid in managing the proof. 

\section{Related Work}
\label{sec:related}

We have designed the \tlaplus proof system as a platform for interactively
verifying concurrent and distributed algorithms. Unlike
most interactive proof assistants~\cite{wiedijk:provers}, TLAPS has been
designed around a declarative proof language that is independent of any specific proof
back-end. \tlaplus proofs indicate what facts are needed to prove a certain
result, but they do not specify precisely how the back-end provers should use
these facts. 
Although this lack of fine control can frustrate users who are intimately
familiar with the inner workings of a particular prover,
declarative proofs are less dependent on specific back-end provers and
less sensitive to changes in their implementation.

We write complex proofs by hierarchically structuring their logic. 
The graphical user interface provides commands that support hierarchical proofs
by allowing a user to zoom in on the current context and by supporting
non-linear proof development. Although some other interactive proof systems such
as Mizar~\cite{trybulec:mizar} and Isabelle/Isar~\cite{wenzel:isar} also offer
hierarchical proofs, to the best of our knowledge these systems do not provide
the Toolbox's abilities to use that structure to aid in reading and writing
proofs and to prove individual steps in any order---facilities that we find
crucial in developing and managing large proofs. The only other proof
assistant that we know to offer a mechanism comparable to our fingerprinting
facility is the KIV system~\cite{balser:kiv}.

The Rodin toolset supporting the Event-B formal method~\cite{abrial:rodin}
shares several aspects with TLAPS: 
Event-B and \tlaplus are both based on set theory, both
emphasize refinement as a way to structure formal developments, and Rodin and
TLAPS mechanize proofs of safety properties with the help of different back-end
provers. Unlike with Event-B models, the structure of \tlaplus specifications is
not fixed: any \tlaplus formula can be considered as a system specification or a
property, and TLAPS does not impose a structure on invariant or refinement
proofs.

Provers designed for program verification such as VCC~\cite{cohen:vcc} or
Why~\cite{herms:certified} target low-level source code rather
than high-level specifications of algorithms. They are based on generators of
verification conditions corresponing to programming constructs, that are
discharged by invoking powerful automatic provers. User interaction is
essentially restricted to the choice of suitable program annotations.

\section{Conclusion}
\label{sec:conclusion}

Using the example of Peterson's algorithm, we have presented the main constructs
of the \tlaplus proof language and, by extension, the ideas underlying the
language design.
That algorithm was chosen because it is well known and simple---so simple that
we had to eschew the use of the SMT solver back-end so we could write a
nontrivial proof. We explained in Section~\ref{sec:real-proofs} why \tlaplus proofs scale
to more complex algorithms and specifications that we do not expect any prover to
handle automatically.
The hierarchical structure of the proof language is essential for giving users
flexibility in designing their proof structure, and it ensures that individual
proof steps are independent of one another. The fingerprinting mechanism of
TLAPS makes use of this independence by storing previously proved results and
retrieving them, even when they appear in a different context.

While not illustrating  the entire proof language~\cite{lamport:tla+2},
Peterson's algorithm does show its main features. Steps correspond to
natural-deduction sequents.
Leaf proofs immediately prove a step, citing the necessary
definitions, facts, and assumptions. Non-leaf proofs consist of another level of
proof steps that end with a \QED step.
This basic structure is oriented towards forward-style
proofs, but the judicious use of backward chaining (\textsc{suffices} steps)
can make proofs more readable. Some features of the proof language
that do not appear in the proof of Figure~\ref{fig:complete-proof} are constructs
for providing a witness to prove an existentially quantified formula,
introducing local definitions, and specifying facts that can be used by the
back-end provers even when they are not explicitly mentioned.

Different proof techniques, such as resolution, tableau methods,
rewriting, and SMT solving offer complementary strengths.
Future versions of TLAPS will probably add new back-end provers.
Adding a new back-end mainly involves writing a translation from
\tlaplus to the input language of the prover.  Such
translations can be complex, and there is a legitimate concern about
their soundness as well as about the soundness of the back-ends
themselves.  For back-ends that can produce proof traces, TLAPS
provides the option to certify the traces within Isabelle.
Proof trace certification has been implemented for Zenon,
and we plan to implement it for other back-end provers including SMT
solvers.  Still, it is much more likely that a proof is meaningless
because of an error in the formula we are proving than because of
an error in a back-end.  Soundness also depends on 
parts of the proof manager.
Users who do not trust its fingerprinting mechanism can
disable it and reprove the entire proof or any part of it.
The proof manager also carries out some critical transformations, such
as replacing $(a+b)'$ by $(a'+b')$.

We cannot overstate how important it is that TLAPS is integrated with the other
\tlaplus tools---especially the TLC model checker. Checking putative invariants
and assertions with TLC on finite instances of a specification is much more
productive than discovering errors during the proof. Users check the exact same
specifications that appear in their proofs. Less obvious is how useful it is
that TLC can evaluate \tlaplus formulas. When verifying a system, we don't want
to prove well-known mathematical facts; we want to assume them. However, it is
easy to make a mistake in formalizing even simple mathematics, and assuming the
truth of an incorrect formula can lead to an incorrect proof. TLC can usually
check the exact \tlaplus formulas assumed in a proof for a large enough instance
to make us confident that our formalization of a correct mathematical result is
indeed correct.

We are actively developing TLAPS. 
The current version supports reasoning about non-temporal formulas, which is
enough for proving safety properties, including invariants and step simulation.
Non-trivial temporal reasoning is required for proving liveness properties, and
our main short-term objective is to support temporal reasoning in TLAPS. It is
not obvious how best to extend natural deduction to temporal logic. We have
designed an approach involving two forms of sequents, expressed with two forms
of the \textsc{assume}/\textsc{prove} statement having
different semantics, that we think will work well.
We also plan to improve support for standard \tlaplus data structures such as
sequences.

%
\makeatletter
\newcommand{\realslash}{/}
\begingroup
\catcode`\/\active
\catcode`\.\active
\catcode`:\active
\gdef\urlslash{\@ifnextchar/{\doubleslash}{\discretionary{}{}{}\realslash}}
\gdef\urlend#1{\let/\urlslash\let.\urldot
                 \discretionary{}{}{}#1\discretionary{}{}{}\endgroup}
\endgroup
\newcommand{\urldot}{.\discretionary{}{}{}}
\renewcommand{\url}{\begingroup\urlbegin}
\newcommand{\urlbegin}{
                       \catcode`\~12\relax
                       \catcode`\#12\relax
                       \catcode`\$12\relax
                       \catcode`\&12\relax
                       \catcode`\_12\relax
                       \catcode`\^12\relax
                       \catcode`\\12\relax
                       \catcode`\/\active
                       \catcode`\.\active
                       \tt
                       \urlend}
\newcommand{\doubleslash}[1]{\discretionary{}{}{}//}

\makeatother 

\bibliographystyle{abbrv}
\bibliography{submission}

\begin{thebibliography}{10}

\bibitem{abrial:rodin}
J.-R. Abrial, M.~Butler, S.~Hallerstede, T.~S. Hoang, F.~Mehta, and L.~Voisin.
\newblock Rodin: an open toolset for modelling and reasoning in {Event-B}.
\newblock {\em STTT}, 12(6):447--466, 2010.

\bibitem{balser:kiv}
M.~Balser, W.~Reif, G.~Schellhorn, K.~Stenzel, and A.~Thums.
\newblock Formal system development with {KIV}.
\newblock In T.~Maibaum, editor, {\em FASE}, volume 1783 of {\em LNCS}, pages
  363--366, Berlin, Germany, 2000. Springer.

\bibitem{smtlib}
C.~Barrett, L.~de~Moura, S.~Ranise, A.~Stump, and C.~Tinelli.
\newblock The {SMT-LIB} initiative and the rise of {SMT}.
\newblock In S.~Barner, I.~Harris, D.~Kroening, and O.~Raz, editors, {\em
  Hardware and Software: Verification and Testing}, volume 6504 of {\em LNCS},
  pages 3--3. Springer, 2011.

\bibitem{bonichon07lpar}
R.~Bonichon, D.~Delahaye, and D.~Doligez.
\newblock Zenon : An extensible automated theorem prover producing checkable
  proofs.
\newblock In N.~Dershowitz and A.~Voronkov, editors, {\em LPAR}, volume 4790 of
  {\em LNCS}, pages 151--165, Yerevan, Armenia, 2007. Springer.

\bibitem{chaudhuri:tlaps}
K.~Chaudhuri, D.~Doligez, L.~Lamport, and S.~Merz.
\newblock Verifying safety properties with the {TLA}\textsuperscript{+} proof
  system.
\newblock In J.~Giesl and R.~H{\"a}hnle, editors, {\em IJCAR}, volume 6173 of
  {\em LNCS}, pages 142--148, Edinburgh, UK, 2010. Springer.

\bibitem{cohen:vcc}
E.~Cohen, M.~Dahlweid, M.~Hillebrand, D.~Leinenbach, M.~Moskal, T.~Santen,
  W.~Schulte, and S.~Tobies.
\newblock {VCC}: A practical system for verifying {Concurrent C}.
\newblock In {\em 22nd Intl. Conf. Theorem Proving in Higher Order Logics
  (TPHOLs 2009)}, volume 5674 of {\em LNCS}, pages 23--42, Munich, Germany,
  2009. Springer.

\bibitem{cousineau:tla-proofs-fm}
D.~Cousineau, D.~Doligez, L.~Lamport, S.~Merz, D.~Ricketts, and H.~Vanzetto.
\newblock {TLA\textsuperscript{+}} proofs.
\newblock In D.~Giannakopoulou and D.~M{\'e}ry, editors, {\em Formal Methods
  (FM 2012)}, volume 7436 of {\em LNCS}, pages 147--154, Paris, France, 2012.
  Springer.

\bibitem{deBruijn72}
N.~G. de~Bruijn.
\newblock {Lambda calculus notation with nameless dummies. A tool for automatic
  formula manipulation with application to the Church-Rosser theorem}.
\newblock {\em Indagationes Mathematicae}, 34:381--392, 1972.

\bibitem{herms:certified}
P.~Herms, C.~March{\'e}, and B.~Monate.
\newblock A certified multi-prover verification condition generator.
\newblock In R.~Joshi, P.~M{\"u}ller, and A.~Podelski, editors, {\em 4th Intl.
  Conf. Verified Software: Theories, Tools, Experiments (VSTTE 2012)}, volume
  7152 of {\em LNCS}, pages 2--17, Philadelphia, PA, 2012. Springer.

\bibitem{lamport:what-good}
L.~Lamport.
\newblock What good is temporal logic?
\newblock In R.~E.~A. Mason, editor, {\em Information Processing 83}, pages
  657--668, Paris, Sept. 1983. IFIP, North-Holland.

\bibitem{lamport03tla}
L.~Lamport.
\newblock {\em Specifying Systems: The {\tlaplus} Language and Tools for
  Hardware and Software Engineers}.
\newblock Addison-Wesley, 2003.

\bibitem{lamport:pluscal}
L.~Lamport.
\newblock The {PlusCal} algorithm language.
\newblock In M.~Leucker and C.~Morgan, editors, {\em ICTAC}, volume 5684 of
  {\em LNCS}, pages 36--60, Kuala Lumpur, Malaysia, 2009. Springer.

\bibitem{lamport:tla+2}
L.~Lamport.
\newblock {\tlatwo}: {A} preliminary guide.
\newblock Draft manuscript, Nov. 2011.
\newblock
  \\\url{http://research.microsoft.com/users/lamport/tla/tla2-guide.pdf}.

\bibitem{lamport:byzantine-paxos}
L.~Lamport.
\newblock Byzantizing {Paxos} by refinement.
\newblock Available at
  \url{http://research.microsoft.com/en-us/um/people/lamport/pubs/web-byzpaxos.pdf},
  2011.

\bibitem{lamport:howtoprove-21century}
L.~Lamport.
\newblock How to write a 21st century proof.
\newblock {\em Journal of Fixed Point Theory and Applications}, Mar. 2012.
\newblock DOI: 10.1007/s11784-012-0071-6.

\bibitem{lu:pastry}
T.~Lu, S.~Merz, and C.~Weidenbach.
\newblock Towards verification of the {Pastry} protocol using
  {TLA\textsuperscript{+}}.
\newblock In R.~Bruni and J.~Dingel, editors, {\em FORTE}, volume 6722 of {\em
  LNCS}, pages 244--258, Reykjavik, Iceland, 2011. Springer.

\bibitem{merz:smt-tlaps}
S.~Merz and H.~Vanzetto.
\newblock Automatic verification of {\tlaplus} proof obligations with {SMT}
  solvers.
\newblock In N.~Bj{\o}rner and A.~Voronkov, editors, {\em LPAR}, volume 7180 of
  {\em LNCS}, pages 289--303, M{\'e}rida, Venezuela, 2012. Springer.

\bibitem{parno:memoir}
B.~Parno, J.~R. Lorch, J.~R. Douceur, J.~Mickens, and J.~M. McCune.
\newblock Memoir: Practical state continuity for protected modules.
\newblock In {\em Security and Privacy}, pages 379--394. IEEE, 2011.

\bibitem{peterson:myths}
G.~L. Peterson.
\newblock Myths about the mutual exclusion problem.
\newblock {\em Inf. Process. Lett.}, 12(3):115--116, 1981.

\bibitem{tlaps}
{The TLAPS Project}.
\newblock Web page.
\newblock \url{http://msr-inria.inria.fr/~doligez/tlaps/}.

\bibitem{trybulec:mizar}
A.~Trybulec.
\newblock Mizar.
\newblock In Wiedijk \cite{wiedijk:provers}, pages 20--23.

\bibitem{wenzel:isar}
M.~Wenzel and L.~C. Paulson.
\newblock {I}sabelle/{I}sar.
\newblock In Wiedijk \cite{wiedijk:provers}, pages 41--49.

\bibitem{wenzel:isabelle}
M.~Wenzel, L.~C. Paulson, and T.~Nipkow.
\newblock The {I}sabelle framework.
\newblock In O.~A. Mohamed, C.~Mu{\~n}oz, and S.~Tahar, editors, {\em TPHOLs},
  volume 5170 of {\em LNCS}, pages 33--38, Montreal, Canada, 2008. Springer.

\bibitem{wiedijk:provers}
F.~Wiedijk, editor.
\newblock {\em The Seventeen Provers of the World}, volume 3600 of {\em LNCS}.
\newblock Springer, 2006.

\end{thebibliography}

\end{document}